\documentstyle[epsf,aps]{revtex}
\begin{document}

\title{Statistical Mechanics of Online Learning of Drifting Concepts
: A Variational Approach.}

\author{Renato Vicente\thanks{rvicente@if.usp.br}}
\address{Instituto de F{\'\i}sica, Universidade de S\~ao Paulo, CP66318,
  CEP 05315-970, S\~ao Paulo, SP Brazil}
\author{Osame Kinouchi \thanks{osame@ultra3000.ifqsc.sc.usp.br}}
\address{Instituto de F{\'\i}sica de S\~ao Carlos, Universidade de S\~ao
  Paulo, CP 369, CEP 13560-970, S\~ao Paulo, SP Brazil}
\author{Nestor Caticha\thanks{nestor@if.usp.br}}
\address{Instituto de F{\'\i}sica, Universidade de S\~ao Paulo, CP66318,
 CEP 05315-970, S\~ao Paulo, SP Brazil}

\maketitle

\abstract{We review the application of Statistical Mechanics methods 
to the study of online learning of a drifting concept in the 
limit of large systems. The model where a feed-forward network learns
from examples generated by a time dependent teacher of the same
architecture  is analyzed. The best possible generalization ability 
is determined exactly, through the use of a variational method. The
constructive variational method also suggests a learning algorithm.  
It depends, however, on some unavailable quantities, such as the 
present performance of the student. The construction of estimators 
for these quantities permits the implementation of a very effective, 
highly adaptive algorithm. Several other algorithms are also studied 
for comparison with the optimal bound and the adaptive algorithm, for
different types of time evolution of the rule. }

\section{Introduction}
The importance of universal bounds to generalization errors, in the spirit of the Vapnik-Chervonenkis (VC) theory, cannot be overstated, since these results are independent of target function and input distribution. This bounds are tight in the sense that for a particular target an input distribution can be found where generalization is as difficult as the VC bound states. However for several learning problems, by making specific assumptions, it is possible to go further. Haussler {\it et al} \cite{haussleretal} have found tighter bounds that even capture functional properties of learning curves, such as for example the occurrence of discontinuous jumps in learning curves, which cannot be predicted from VC theory alone. 

These results were derived by adapting to the problem of learning, ideas that arise in the context of Statistical Mechanics. In recent years many other results \cite{seung,watkin,kinzel}, bounds or approximations, rigorous or not, have been obtained in the learning theory of neural networks by applying a host of methods originated in the study of disordered materials. These methods permit 
looking at the properties of large networks, where great analytical 
simplifications can occur; and also, they afford the possibility of 
performing averages over the randomness introduced by the training 
data. They are useful in that they give information about typical rather 
than e.g. worst case behavior and should be regarded as complementary
to those of Computational Learning Theory.

The Statistical Mechanics of learning has been formulated either as a 
problem at thermodynamical equilibrium or as a dynamical process 
off-equilibrium, depending on the type of learning strategy. Although many 
intermediate regimes can be identified, we briefly discuss the two 
dynamical extremes. Batch or offline methods 
essentially give rise to the equilibrium formulation, while online 
learning can be better described as an off equilibrium process.

The offline method begins by constructing a cost or energy function 
on the parameter space, which depends on {\it all the training data} 
simultaneously \cite{seung,watkin}. Learning occurs by defining a 
gradient descent process on the parameter space, subject to some 
(thermal) noise process, which permits to some extent escaping from 
local traps. In a very simplified way it may be said that, after some time, this process leads to ``Thermal equilibrium'', when essentially all possible information has been extracted by the algorithm from the learning set. The system is now described by a stationary (Gibbs) probability distribution  on parameter space.

On the other extreme lies online or incremental learning. Instead of training
 with a cost function defined over all the available examples, the
 online cost function depends directly on only {\it one single example}, independently chosen at each time step of the learning 
 dynamics \cite{amari67},(for a review, \cite{coolen97}). 
 Online learning occurs also by gradient descent, but now the random nature of the presentation of the examples implies that at each learning step 
 an effectively different cost function is being used. This 
can lead to good performances even without the costly memory resources needed to keep the information about the whole learning set, as is the case in the offline case. 

Although most of the work has concentrated on learning in stationary 
environments with either online or offline strategies, the learning of 
drifting concepts has also been modeled using ideas of Statistical 
Mechanics \cite{biehl92,biehl93,kc93}. The natural approach to this 
type of problem is to consider online learning, since old examples 
may not be representative of the present state of the concept. 
It makes little sense, if any, to come to thermal equilibrium with 
possibly already irrelevant old data, as would be the case of an 
offline strategy. The possibility of forgetting old information and 
of preventing the system from reusing it,  
which are essential features to obtain good performance, 
are inherent to the online processes we will see below.

We will model the problem of supervised learning, 
in the sense of Valiant \cite{valiant}, of a drifting concept by defining a
 ``teacher'' neural network. Drift is modeled by allowing 
the teacher network parameters to undergo a drift that can be either 
random or deterministic. The dynamics of learning occurs in discrete time. 
At each time step, a random input vector is chosen independently 
from a distribution $P_D$, giving rise to a temporal stream of input-output 
pairs, where the output is determined by the teacher. From this set of data 
the student parameters will be built.

The question addressed 
in this paper concerns the {\it best} possible way in which the 
information can be used by the student in order to obtain maximum 
typical generalization ability. This is certainly too much to ask for 
and we will have to make some restrictions to the problem.
This question will be answered by means of a {\it variational
method} for the following class of exactly soluble models: 
a feed-forward boolean
network learning from a teacher which is itself a neural network of 
similar architecture and learns by a hebbian modulated mechanism. 

This is still hard and further restrictions will be made.
The thermodynamic limit (TL) will always 
be assumed. This means that the dimension $N$ 
of parameter space is taken to infinity. 
This increase brings about great analytical simplifications. 
The TL is the natural regime to study in Condensed Matter 
Physics. There the number of interacting units is of the order of 
$N\approx 10^{23}$ and fluctuations of macroscopic variables of order
$\sqrt{N}.$ In studying neural networks, results obtained in the TL
ought to be considered as the first term in a systematic expansion in 
powers of $1/N.$  

Once this question has been answered in a restricted setting, what does it 
imply for more general and realistic problems? The variational method
has been applied to several models, including boolean and soft transfer 
functions, single layer perceptrons and networks with hidden units, 
networks with or without overlapping receptive fields and also for 
the case of non-monotonic transfer functions 
\cite{kc92b,kc95,cc95,sc96,vc97,vandenbroeck}. 
In solving the problem in 
different cases, different 
optimal algorithms have been found. But rather than delving in the 
differences, it is important to stress that a set of features 
is common to all optimal algorithms. Some of these common features are 
obvious or at least expected and have been incorporated into algorithms 
built in an {\it ad hoc}  manner. Nevertheless, it is quite interesting 
to see them arise from theoretical arguments rather than heuristically.
Moreover, the exact functional dependence is also obtained, and this can 
never be obtained just from heuristics. See \cite{opper} for an explicitly 
Bayesian formulation of online learning which, in the TL seems to be similar 
to the variational method. 

The first important result of the variational
program is to give lower bounds to the generalization errors. But it gives
more, the constructive nature of the method furnishes also an `optimal 
algorithm'. However, the direct implementation
of the optimal algorithm is not possible, as it relies on the knowledge
of information that is not readily accessible. 
This reliance is not to be thought of as a
drawback but rather as indicating what kind of information is needed 
in order to approximate, if not saturate, the optimal bounds. 
It indicates directions for further research where the aim should be on 
developing efficient estimation schemes for those variables.

The procedure to answer what is the best possible algorithm in the sense of
generalization is as follows. The generalization error, in the TL, can be
written as a function of a set of macroscopic parameters, sometimes referred to as `order parameters', by borrowing the nomenclature from Physics.  
The online dynamics of the weights (microscopic
variables) induces a dynamics of the order parameters, which in the 
TL is described by a {\bf closed set} of coupled differential equations. 
The evolution of the generalization error is thus a functional of the 
cost function gradient which defines the learning algorithm. The gradient 
of the cost function is usually called the modulation function. The local 
optimization (see \cite{saadglobal} for global) is done in the following way. Taking the functional derivative of the rate of decay of the generalization error, with respect to the modulation function, equal to zero, permits determining the modulation function that extremizes the mean decay at each time step. This extreme represents, in many of the interesting cases a maximum ( see \cite{vc97} for exceptions).
We can thus determine the modulation function, i.e. the algorithm, that leads to the fastest local decrease of the generalization error under several restrictions, to be discussed below. 

In this paper several online algorithms are
analyzed for the boolean single layer perceptron. Other
architectures, with e.g. internal layers of hidden units, can be analyzed,
although there is a need for laborious modifications of the methods. Examples of random 
drift, deterministic evolution, changing drift levels 
and piecewise constant concepts are presented.
The paper is organized as follows. In
section II, the variational approach is briefly reviewed. In section 
III, analytical results and simulations are presented for several 
algorithms in the cases of random  drift and 
deterministic `worst-case' drift, where the teacher flees from the student,
in weight space. The asymptotics of the different algorithms are 
characterized by a couple of exponents, $\beta$, the learning exponent and 
$\delta$, the drift or residual exponent. 
A relation between these exponents is obtained.
A practical adaptive algorithm is discussed in section IV, where
it is applied to a problem with changing drift level. In section V,
the Wisconsin test for perceptrons is studied. 
Numerical results for the piecewise constant rule are presented. 
Concluding remarks are presented in section VI.

\section{The variational approach}

The mathematical framework employed in the statistical mechanics of
online learning  and in the variational optimization are quickly reviewed 
in this section. We consider only the simple perceptron with no hidden
layer. For extensions to other architectures see
\cite{kc95,cc95,sc96,vc97,vandenbroeck}


\subsection {Preliminary definitions}

The boolean single layer perceptron is defined by the  function
$\sigma_B=sign({\bf B}\cdot {\bf S})$, with ${\bf S}\in {\cal R}^N$,
parametrized by the {\it concept} weight vector  ${\bf B}\in{\cal R}^N$  also 
called  {\it synaptic vector}.
 
In the student-teacher scenario that we are considering, 
a perceptron ({\it teacher}) generates a
sequence of statistically independent training pairs ${\cal L}=
\{({\bf S}^\mu,{\sigma}_B^\mu):\mu=1,...,p\}$, and
another perceptron ({\it student}) is constructed, using only  the examples 
in ${\cal L}$, in order to infer the concept 
represented by the teacher's vector.
The teacher and student are respectively defined by weight vectors
${\bf B}$ and ${\bf J}$ with  norms  denoted by $B$ and $J$.

In the presence of noise, instead of 
$\sigma_B$, the student has access only to  
a corrupted version $\tilde{\sigma}_B$.
For example, for {\it multiplicative} noise, each teacher's output 
is flipped independently 
with probability $\chi$ \cite{biehl95,ckk,copelli97,Heskes}: 
\begin{equation}
P(\tilde{\sigma}_B|\sigma_B)
=(1-\chi)\delta(\sigma_B,\tilde{\sigma}_B)+\chi\delta(\sigma_B,-\tilde{\sigma}_B)\:,
\label{eq:noise}
\end{equation}
where  $\sigma_B=sign(y)$, and $y={\bf B \cdot S}/B$ is the
{\it normalized field}. The Kronecker $\delta$ is $1 \; (0)$
only if the arguments are equal (different). In the same way,  for the
student, the field  $x={\bf J \cdot S}/J$ and the output
$\sigma_J=sign(x)$ are defined.

The definition of a global cost function
$E_{\cal L}({\bf J })=\sum_\mu E^\mu({\bf J})$, over the entire data set 
${\cal L}$, is required for batch or offline learning.
The interaction among the partial potentials $E^\mu({\bf J})$ may
generate spurious local 
minima, leading to metastable states and possibly very long 
thermalization times. This can be avoided to a great extent by 
learning online.  

We define a discrete dynamics where at 
each time step a weight update 
is performed along the gradient of a partial potential $E^\mu({\bf J })$, 
which depends on the randomly chosen  $\mu^{th}$ example.
This random sampling of partial potentials introduces fluctuations which
tend to decrease as the system approaches a (hopefully) global minimum. 
That  process has been recently 
called {\it self-annealing} \cite{Hondou} in opposition to the external 
parameter dependent simulated annealing.  
The general conditions for convergence of online learning to a global 
minimum, even in stationary environments, 
is an open problem of major current interest.

The online dynamics can be
represented by a finite difference equation for the update of 
weight vectors. For each new random example, make a small correction of
the current student, in the direction 
opposite to the gradient of the partial potential 
and also allow for a restriction of the overall length of the weight vector
to prevent runaway behavior:

\begin{equation}
{\bf J}^{\mu+1}={\bf J}^{\mu} - \Delta t {\Omega^\mu}{\bf J}^{\mu}
-\Delta t \nabla_{\bf J} E^{\mu} \:.
\label{eq:learn0}
\end{equation} 
Here the partial potential $E^{\mu}$ is a function of the scalars
 that are accessible to the student (field $x$, norm $J$ and output 
 $\tilde{\sigma}_B$) and  corresponds to the randomly sampled example pair 
$({\bf S}^\mu, {\tilde \sigma}_B^\mu)$. The time scale $\Delta t$ must be
$\Delta t \sim {\cal O}(1/N)$ so that 
in the TL we can derive well-behaved differential equations,  
in general we will choose $\Delta t = 1/N $. The second term 
allows to control the norm of the weight vector {\bf J}.

It is simple to see that  $\nabla_{\bf J} E^{\mu}=
(\partial E^{\mu}/\partial x) \nabla_{\bf J} x$. The calculation of the 
 gradient and the definition $\partial E^{\mu}/\partial x=
 -J^{\mu}W^{\mu}({\cal V})\tilde{\sigma}_B^\mu$ finally lead to the online
 dynamics in the form:

\begin{equation}
{\bf J}^{\mu+1}=\left(1-\frac{\Omega^\mu}{N}\right){\bf J}^{\mu}
+\frac{1}{N}J^{\mu}W^{\mu}({\cal V})\tilde{\sigma}_B^\mu {\bf S}^{\mu}\:.
\label{eq:learn}
\end{equation} 
Note that each example pair $({\bf S}^\mu,\tilde{{\sigma}}_B^\mu)$ is
used only once to update the student's synapses, 
$\tilde{\sigma}_B^\mu {\bf S}^{\mu}$ is called the {\it hebbian term} and
the intensity of each modification is given by the {\it modulation function}
$W$. The prefactor $J \tilde{{\sigma}}_B^\mu$ 
can be absorbed into the modulation function, but
has been explicitly written for convenience. The single most
important fact is that the relevant change is made along the direction 
of the input vector ${\bf S}^\mu$. The reasons for this restriction are the 
following: the optimal algorithms to be discussed bellow are bounds only 
on the TL. In the absence of site correlations, i.e. 
$\langle S_i S_j \rangle=0$,   
the prefactor of ${\bf S}^\mu$ is a diagonal matrix \cite{opper}. 
Furthermore the class of modulated hebbian algorithms are interesting even 
for finite $N$ from a biological perspective. 
The symbol ${\cal V}$ denotes
the {\it learning situation}, that is, the set of quantities that 
we are allowed to use in the modulation function, that is, the available 
information. 
For boolean perceptrons, ${\cal V}$ may contain the corrupted
teacher's output $\tilde{\sigma}_B$, the field $x$, and as 
discussed below, some information about the generalization error. We
can still study the restrictions ${\cal V}=\{\tilde{\sigma}_B\}$, 
${\cal V}=\{\tilde{\sigma}_B,\sigma_J\}$ and 
${\cal V}=\{\tilde{\sigma}_B,\mid x\mid\}$. Evidently, the more information
the student has, the better we expect it to learn.

We consider a specific model for concept drift introduced by 
Biehl and Schwarze \cite{biehl92}. The drift scale that can be followed
by an online learning system can not be to large for it would 
be impossible to track, but if too slow it trivially reduces to an 
effectively stationary problem. Their choice, which makes the problem interesting,
is as follows.
At each time step the concept vector ${\bf B}$ suffers the influence of the
changing environment and evolves as

\begin{equation}
{\bf B}^{\mu+1}=\left(1-\frac{\Lambda^\mu}{N}\right){\bf B}^{\mu} +
\frac{1}{N}\vec{\eta}^{\mu}\:,
\label{eq:drift}
\end{equation}
where $\Lambda$ controls the norm $B$ and $\vec{\eta}\in {\cal R}^N$
is the {\it drift vector}. 
Random and deterministic versions of $\vec{\eta}$ will be considered 
in section 4.

The performance of a specific student ${\bf J}$ in a given concept ${\bf B}$
can be measured by the generalization error $e_G$ that is defined as the
instantaneous average error $\epsilon=\frac{1}{2}(1-\sigma_J\sigma_B)$
($\sigma_B$ is the non-corrupted output) over inputs extracted
from the uniform distribution $P_{\cal U}({\bf S})$ 
with support over the hyper-sphere of radius
$\sqrt{N}$:

\begin{equation}
e_G({\bf J},{\bf B})=\int d{\bf S} P_{\cal U}({\bf S})\;
\epsilon\left({\sigma_J}({\bf S}),{\sigma_B}({\bf S})\right)\:.
\label{eq:eg_def}
\end{equation}

We make explicit the difference between 
$e_G$ and the prediction error $e_P$, which measures the average  
$\epsilon_P=\langle\frac{1}{2}(1-\sigma_J\tilde\sigma_B)\rangle$, 
over the {\it true} distribution of examples $P_D$. 
It is not difficult to see that
the expression for $e_G$ is invariant under  rotations of axes
in ${\cal R}^N$, therefore the integral (\ref{eq:eg_def}) depends only
on the invariants $\rho={\bf B\cdot J}/BJ$, $x$, $y$, $B$ and $J$.
In the TL a straightforward application of the Central Limit Theorem
leads to:
\begin{eqnarray}
\nonumber e_G(\rho)&=&\int dx dy P_{\bf
                   C}(x,y)\;\frac{\Theta(-xy)}{2}\\
                   &=& \frac{1}{\pi}\arccos{\rho} \:.
\label{eq:eg_clt}
\end{eqnarray}
$P_{\bf C}(x,y)$ is a Gaussian distribution in ${\cal R}^2$ with
correlation matrix
\[{\bf C}=\left( \begin{array}{cc}
                        1&\rho\\
                        \rho&1
                  \end{array}\right) \:.\]

Note that  $\rho$ is a  parameter
in the probability distribution describing the fields, and $J$ and $B$
define the scale of the same fields. In statistical physics, 
that variables are called  {\it macroscopic variables}.

\begin{figure}
\begin{center}
\leavevmode
\epsfxsize =.2\textwidth
\epsfbox{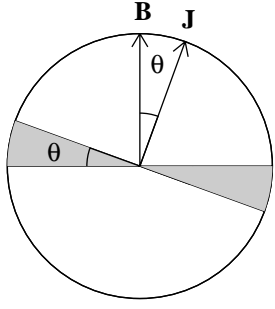}
\caption{Simple representation  of  weight vectors in  the
  hyper-sphere. The teacher and the student disagree when the input
  vector {\bf S}
  is inside the shaded region.} 
\label{fig1}
\end{center}
\end{figure}

The intuitive
meanings of $\rho$ and of  the Eq.  
\ref{eq:eg_clt}  can be  verified with the help of   
Fig. \ref{fig1},  observing that  $\rho = \cos \theta$ and that for
the boolean perceptron  the weight vectors are  normal  to a
hyper-plane that divides the hyper-sphere in two differently
labelled hemispheres,  it is easy to see that  
the student and the professor disagree on the  labeling of input
vectors ${\bf S}$ inside the shaded region, thus trivially
$e_G=\theta/\pi=\frac 1 \pi \arccos \rho$  . 
 
\subsection{Emergence of the  macroscopic dynamics}  

The dimensionality of the dynamical system involved can be reduced by
using (\ref{eq:drift}) and (\ref{eq:learn}) to 
write a system of coupled difference equations to the macroscopic variables:
\begin{eqnarray}
\nonumber \rho^{\mu+1}&=& \rho^{\mu} + \frac{\rho^{\mu}}{N}\left[
W^{\mu}({\cal
V})\left(\frac{y^{\mu}\tilde{\sigma}^{\mu}_B}{\rho^{\mu}}-
\Delta^{\mu}\right)-
\frac{1}{2}(W^{\mu}({\cal V}))^{2}\right] \\
&+&\frac{1}{N}\left[\frac{{\bf
J}^{\mu}\cdot{\vec{\eta}}^{\mu}}{J^\mu}-
\rho^{\mu}\Lambda^\mu +\frac{ {\bf S}^{\mu}\cdot 
\vec{\eta}^\mu}{N}\tilde{\sigma}^{\mu}_BW^{\mu}({\cal V})\right]+
{\cal O}\left(\frac{1}{N^2}\right)
\label{eq:dif1} \:,\\
 J^{\mu+1}&=&J^{\mu}+\frac{J^{\mu}}{N}\left(W^{\mu}({\cal V})  
\Delta^{\mu} +\frac{1}{2}(W^{\mu}({\cal V}))^2 -\Omega^{\mu}   
\right)+{\cal O}\left(\frac{1}{N^2}\right)
\label{eq:dif2} \:, \\
 B^{\mu+1}&=&B^{\mu}+\frac{1}{N}\left(\frac{\vec{\eta}^\mu\cdot{\bf B}^\mu}{B} 
-\Lambda^{\mu}B^{\mu}\right)+\frac{\vec{\eta}^{\mu}\cdot
\vec{\eta}^\mu}{2BN^2} + {\cal O}\left(\frac{1}{N^2}\right)
\label{eq:dif3} \:.
\end{eqnarray}

In the above equations the {\it local  stability} $\Delta=x\tilde{\sigma}_B$ was introduced. 
Positive stability means that the student classification $\sigma_J=sign(x)$ agrees with
the (noisy) learning data $\tilde{\sigma}_B$. 
 
The usefulness of the TL lies in the possibility of transforming the
stochastic difference equations into a closed set of deterministic 
differential equations \cite{rujan,kc92a}. The idea is 
to choose a continuous time scale $\alpha$ such that for the TL regime
$p/N \rightarrow \alpha$, where $p$ is  the number of 
examples already presented. The equations are then averaged over the
input vectors $\bf S$ and drift
vector $\vec{\eta}$ distributions, leading to:  
\begin{eqnarray}
\frac{d\rho}{d\alpha}&=&\rho\left\langle  W
\left(\frac{y\tilde{\sigma}_B}{\rho}-\Delta\right)-
\frac{1}{2}W^{2}\right\rangle+
\left\langle\frac{{\bf J}\cdot{\vec{\eta}}}{J}-
\rho\Lambda + C_{S\eta}\tilde{\sigma}_B 
W\right\rangle
\label{eq:1} \:,\\
\frac{dJ}{d\alpha}&=&J\left\langle W  
\Delta +\frac{1}{2}W^2 -\Omega \right\rangle
\label{eq:2} \:, \\
 \frac{dB}{d\alpha}&=&\left\langle\frac{\vec{\eta}\cdot{\bf B}}{B}-\Lambda
 B+C_{\eta\eta}\right\rangle
\label{eq:3} \:,
\end{eqnarray}
where $\langle...\rangle =\int d{\vec{\eta}} \:d{\bf S}\,
(\ldots) P(\vec{\eta},{\bf S})$ and the definitions
$C_{S\eta}=\lim_{N\rightarrow \infty}({\bf S}
\cdot{\vec{\eta}})/(NB)$ and \mbox{$C_{\eta\eta}=\lim_{N\rightarrow \infty}({\vec{\eta}
\cdot\vec{\eta}})/(2BN)$} have been used.

The fluctuations in the stochastic equations vanish in the TL and the
above equations become exact ({\it self-averaging property}). This can
be proved by writing  the
Fokker-Planck equations for the finite $N$ stochastic process defined
in  (\ref{eq:dif1}), (\ref{eq:dif2}) and (\ref{eq:dif3}), and
showing that the diffusive term vanishes in the TL \cite{coolen97}.

\subsection{Variational optimization of algorithms}

The variational approach was  proposed 
in \cite{kc92b} as an analytical method to find learning
algorithms with optimal mean decay (per example)  of the generalization
error. The same method was applied in several
architectures and learning situations. 

The idea is to write: 
\begin{equation}
\small
\frac{de_G}{d\alpha}=\frac{\partial e_G}{\partial \rho} \frac{d\rho}{d\alpha}\:,
\end{equation}
and use equation (\ref{eq:1})  to build the functional:
\begin{equation}
\frac{de_G}{d\alpha}[W] = \frac{\partial e_G}{\partial \rho}
\left[\rho\left\langle  W
\left(\frac{y\tilde{\sigma}_B}{\rho}-\Delta\right)-
\frac{1}{2}W^{2}\right\rangle+
\left\langle\frac{{\bf J}\cdot{\vec{\eta}}}{J}-
\rho\Lambda + C_{S\eta}\tilde{\sigma}_B W\right\rangle\right]\:.
\end{equation}  
Thus, the optimization is attained by imposing the extremum 
condition:
\begin{equation} 
\frac{\delta}{\delta W({\cal V})}\left( \frac{de_G}{d\alpha}[W({\cal
    V})]\right)_{W=W^*}=0\:,
\end{equation}
 where ${\delta}/{\delta W({\cal V})}$ stands for the functional
derivative in the subspace of modulation functions $W$ with dependence
in the set $\cal V$. The above equation can be solved  observing that 
$\partial e_G/\partial \rho \neq 0$ and 
\begin{equation}
\frac{\delta}{\delta W({\cal V})} \langle f({\cal H},{\cal V})
W^n({\cal V})\rangle =
n\langle  f({\cal H},{\cal V})\rangle_{\cal H |\cal V} \:W^{n-1}({\cal V})\:, 
\end{equation}
 $f$ is an arbitrary function, $\cal H$ is the set of {\it hidden}
 variables, that is, in contrast with the set ${\cal V}$, the variables
 not accecible to the student (e.g. fields
$y$, drift vector $\vec{\eta}$, etc ...) and 
$\langle ...\rangle_{\cal H |\cal V} = \int d{\cal H} P({\cal H}|{\cal
  V}) ...$, where for a given set ${\cal H}=\{a_1,a_2,..\}$,  
$d{\cal H}=da_1da_2 ...$. The solution is given by:
\begin{equation}
W^*({\cal V})=\left \langle\frac{(y+C_{S\eta})\tilde{\sigma}_B}{\rho}-\Delta\right\rangle_{{\cal
  H}|{\cal V}}.
\label {eq:optmodul}
\end{equation}
By writing $y+C_{S\eta}=({\bf B}+\vec{\eta})\cdot{\bf S}/\sqrt{B}$ it
can be seen that the optimal algorithm ``tries'' to pull the example
stability $\Delta$, not to an estimative of the present teacher
stability, but to one already corrected by the effect of the drift 
$\vec{\eta}$. It seems natural to concentrate on cases where the drift 
and the input vectors are uncorrelated ($C_{S\eta}=0$).

The optimization under different conditions of information
availability, i.e., different  specifications of the sets $\cal V$ and
$\cal H$, leads to different learning algorithms. This can be seen by 
performing the appropriate averages in (\ref{eq:optmodul}), as we 
proceed to show: 

\begin{description}
\item[Annealed Hebb Algorithm:]{ Suppose that the available
 information is limited to the corrupted teacher output. This
 corresponds to the learning situation such that ${\cal H}=
 \{y,\vec{\eta},|x|,\sigma_J\}$ and ${\cal V}=\{\tilde{\sigma}_B\}$.
 Considering that $C_{S\eta}=0$, we need to perform the average: 
 \begin{equation}
 W^*(\tilde{\sigma}_B)=\frac{\tilde{\sigma}_B}{\rho}
 \left \langle y-\rho x\right\rangle_{\{x,y\}|\tilde{\sigma}_B},
 \label{eq:calculo} 
 \end{equation}
 which involves $
 \int dy \: y \: P\left(y\mid \tilde{\sigma}_B\right) ; \;
 \int dx \: x \: P\left(x\mid \tilde{\sigma}_B\right).$
 The probability distributions are easily obtained using
 Bayes theorem:
\begin {equation}
P(y\mid \tilde{\sigma}_B)=\frac {P(\tilde{\sigma}_B\mid y)P(y)}
{\int dy \: P(\tilde{\sigma}_B\mid y)}.
\label{eq:distr}
\end{equation}
By the Central Limit Theorem we know that, in the TL, $P(y)$ and $P(x)$
are Gaussians with unit variance and it is not difficult to verify, 
using (\ref {eq:noise}), that :
\begin {equation}
 P(\tilde{\sigma}_B\mid y)=\frac{\chi}{2} + 
 (1-\chi)\Theta(\tilde{\sigma}_B y); \;
 P(\tilde{\sigma}_B\mid x)=\frac{\chi}{2} + 
 (1-\chi)H(-\frac{\tilde{\sigma}_B x}{\lambda}),
\end{equation}
 where $\lambda=\sqrt{1-\rho^2}/\rho$ and 
 $H(x)=\int_x^\infty \frac{dt}{\sqrt {2\pi}}e^{-t^2/2}$.

It follows that
\begin{equation}
\left \langle y\right\rangle_{\{x,y\}|\tilde{\sigma}_B} = 
\sqrt{\frac 2 \pi} \tilde{\sigma}_B (1-\chi); \;\;\;
\left \langle\rho x\right\rangle_{\{x,y\}|\tilde{\sigma}_B} = 
\sqrt{\frac 2 \pi} \tilde{\sigma}_B (1-\chi)\rho^2.
\end{equation}
Combining the above results in (\ref{eq:calculo}) finally gives : 
\begin {equation}
\label {eq:hebb}
W_{AH} (\tilde{\sigma}_B;\rho,\chi)= \sqrt{\frac{2}{\pi}}
\lambda^2 \rho (1-\chi)\:.
\end{equation}

The weight changes are proportional to the Hebb factor, but the
modulation function does not depend on the 
example stability $\Delta$ (see Fig.~\ref{MF}). Hence the name Hebb. However
this function is not constant in time, the temporal evolution  
({\it annealing}) is automatically incorporated into the modulation 
function. Optimal annealing is achieved by having the modulation function 
depend on the parameter $\rho$, the normalized overlap between the student
${\bf J}$ and the concept ${\bf B}$. Since this quantity is certainly not 
available to the student, there will be a need to complement 
the learning algorithm with an efficient estimator of the present level of 
performance by the student (see Sec. IV).

}

\begin{figure}
\epsfxsize =.7\textwidth
\begin{center}
\leavevmode
\epsfbox{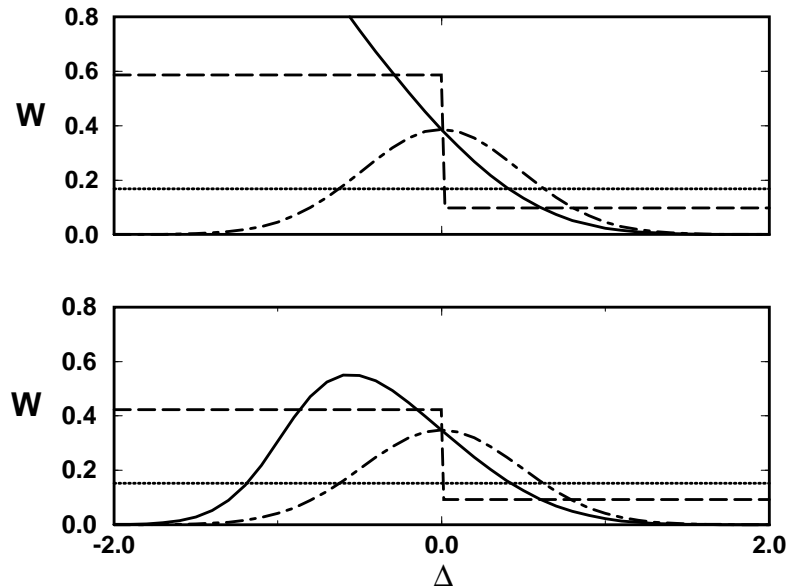}
\caption{Modulation functions exemplified for $\rho=0.9$ 
with noise levels $\chi=0$ (top) and 
$\chi=0.1$ (bottom):
Annealed Hebb (dots), Step (dashes), Symmetric 
(dashes-dots) and Optimal (solid).} 
\label{MF}
\end{center}
\end{figure}
 
\item[Step Algorithm (\cite{copelli97}):]{This algorithm,
obtained under the restriction ${\cal H}=\{y,\vec{\eta},\\|x|\}$ 
and ${\cal V}=\{\tilde{\sigma}_B,\sigma_J\}$, is a close relative
to Rosemblatt's original perceptron algorithm, which works by error correcting
and treats all the errors in the same manner. There are two important
differences, however, 
since correct answers also cause (smaller) corrections and 
furthermore, the size of the corrections evolves in time in a similar manner  
to the annealed Hebb algorithm.

The modulation function is
\begin {equation}
\label {eq:step}
W_{Step} (\tilde{\sigma}_B,\sigma_J;\rho,\chi)=\frac{1}{\sqrt{2\pi}}\lambda^2 \rho (1-\chi)\frac{1}{\left[\frac{\chi}{2}+\frac{(1-\chi)}{\pi}
\arccos(-\rho\tilde{\sigma}_B\sigma_{J})\right]}\:.
\end{equation}
Note that the Step Algorithm has access to the student's output and can
differentiate between right ($\Delta>0$) and 
wrong ($\Delta<0$) classifications,
the name arises from the form of its modulation function 
(Fig.~\ref{MF}). The annealing increases the height of the step, i.e
the difference between right and wrong, 
as the overlap $\rho$ goes to one.

  }
 
\item[Symmetric Weight Algorithm (see \cite{kc92a}):]{ This is the optimal 
algorithm for the learning situation described by ${\cal H}=\{y,\vec{\eta},\sigma_J\}$ and ${\cal V}=
\{\tilde{\sigma}_B,|x|\}$. The resulting modulation function is given by:

\begin {equation}
\label {eq:SW}
W_{SW} (\tilde{\sigma}_B,|x|;\rho,\chi)= \sqrt{\frac
  2{\pi}}\lambda (1-\chi)
e^{-x^2/2\lambda^2}.
\end{equation}

That algorithm cannot discern between wrong and right classifications, but
 only differentiates between ``easy'' (large $\mid\Delta\mid$) and ``hard'' 
(small $\mid\Delta\mid$) classifications, concentrating the learning in
``hard'' examples (Fig.~\ref{MF}).  
}

\item[Optimal  Algorithm (see \cite{kc92b,biehl95,ckk}):]
{When  all the available information is used  we  have the learning
situation described by ${\cal H}=\{y,\vec{\eta}\}$ and ${\cal
      V}=\{\tilde{\sigma}_B,|x|,\sigma_J\}$. the optimal algorithm is then
given by:
          
\begin {equation}
\label {eq:OPT}
W_{OPT} (\Delta=\tilde{\sigma}_B x;\rho,\chi)= \frac{1}{\sqrt{2\pi}}\lambda (1-\chi)
\frac{e^{-\Delta^2/2\lambda^2}}{\chi/2 +(1-\chi)H(-\Delta/\lambda)} \; \;.
\end{equation} 

In the presence of noise, a crossover is built into 
the Optimal modulation function. This crossover is from a regime where the student
classification is not strongly defined ($\Delta$ negative but small) -- and the 
information from the teacher is taken seriously -- 
to a regime where the student is confident on 
its own answer -- and any strong disagreement (very negative $\Delta$) 
with the teacher will be attributed to noise, and thus effectively
disregarded. The scale of the stabilities where 
the crossover occurs depends on
the level of performance $\rho$ and therefore is also annealed.  
} 
\end{description}

The learning mechanisms are highly adaptive and remain the same  
in the case of drifting rules, where 
the common features described above, mainly the $\rho$ dependent annealing, 
lead automatically to a forgetting mechanism without the need 
to impose it, based on heuristic expectations, in an {\it ad hoc} manner. 

It is interesting to note that the heuristically proposed
algorithms are approximations of these optimized modulation functions. For
instance, simple Hebb rule is the Annealed Hebb when
$\vec{\eta}=0$, since it can be shown in this case that $WJ=1$,  and
corresponds to the $\rho\rightarrow 0$ regime of all the optimized 
algorithms;
Rosemblatt Perceptron algorithm is qualitatively similar to the
Step algorithm; Adatron \cite{anlauf,watkin} approximates the Optimal
algorithm for $\chi = 0$ and $\rho \rightarrow 1$;
OLGA \cite{KS} and Thermal perceptron \cite{Frean} algorithms
resemble the Optimal modulation with $\chi>0$.

\section{Learning drifting concepts}

The important result from last section is that, under the
assumption of uncorrelated drift and input vectors, 
the modulation functions do not depend on the drift parameters
(in contrast with the explicit dependence on the
examples noise level $\chi$). So, they are expected to be robust to continuous
or abrupt, random or deterministic, concept changes. In this section
simple instances of drifting concepts  
are examined; abrupt changes are studied in Sec. V. 
 
\subsection{Random drift}

In this scenario, the concept weight vector 
${\bf  B}$ performs a random walk on the
surface of a $N$-dimensional unit sphere. 
The drift vector has random components with zero mean and variance $2D$,
\begin{equation}
\langle \eta_i^\mu \eta_j^\nu \rangle = 2 D \delta_{ij} \delta_{\mu\nu}\:,
\end{equation}
The condition
${\bf B}^{\mu+1} \cdot {\bf B}^{\mu +1} = 1$ is imposed
in (\ref{eq:3}) by considering that ${\bf B}^{0} \cdot {\bf B}^{0} =
1$ and with the choice
$
\Lambda = \vec{\eta}\cdot{\bf B} + D. 
$

\begin{figure}
\epsfxsize =.7\textwidth
\begin{center}
\leavevmode
\epsfbox{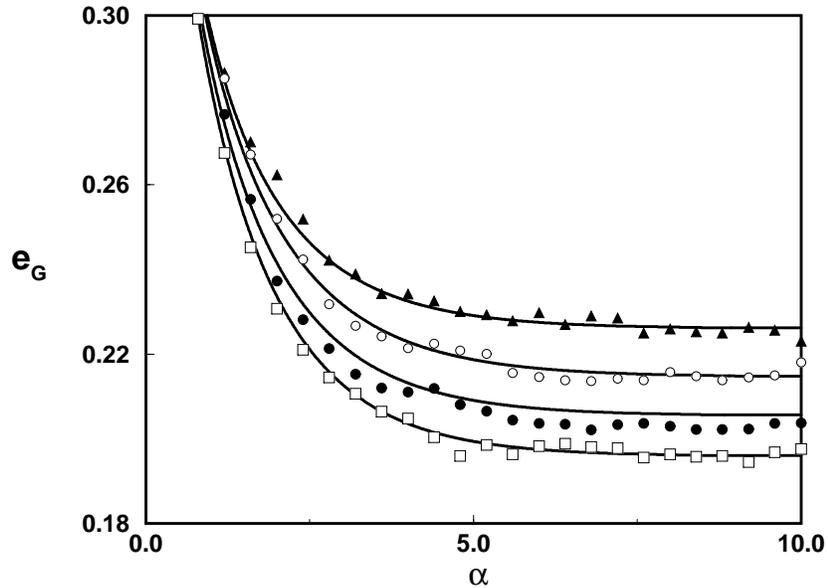}
\caption{Integration of learning equations and simulation results
  ($N=5000$) for random drift $D=0.1$: Annealed Hebb (triangles), 
Symmetric (white circles), Step (black circles) and Optimal (white
squares). The self-averaging property is clear since the simulation
results refer to only one run.} 
\label{eg-random1}
\end{center}
\end{figure}

\begin{figure}
\epsfxsize =.7\textwidth
\begin{center}
\leavevmode
\epsfbox{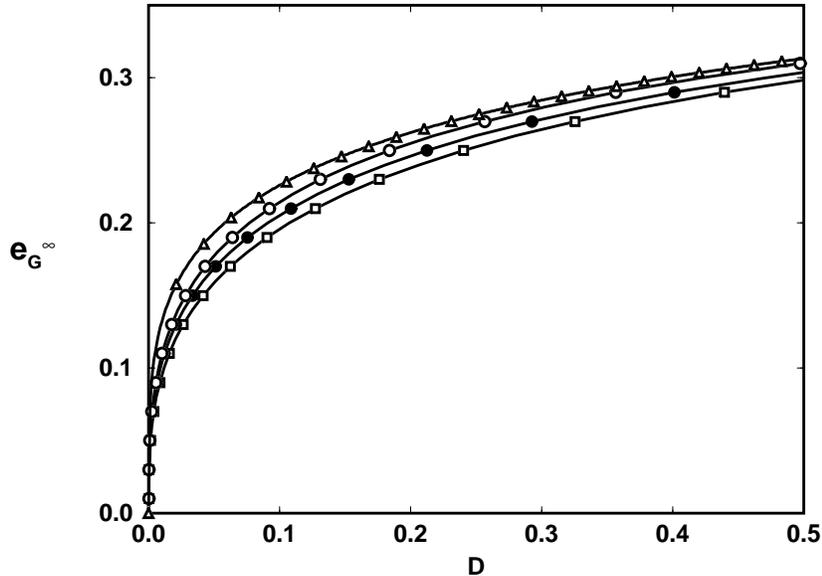}
\caption{Asymptotic error $e_G^\infty(D)$ for random drift: 
Annealed Hebb (triangles),
Symmetric (white circles), Step (black circles) and Optimal
(white squares).} 
\label{eg-random2}
\end{center}
\end{figure}

The order of magnitude of the
scaling of the drift vector with $N$ is important since it
gives the correct time scale for non trivial behavior: if smaller, 
the drift would be
irrelevant in the time scale of learning, while if larger 
it would not allow any tracking. In the relevant regime, the
task in nontrivial also in another sense: the autocorrelation of the
concept vector decays exponentially in the $\alpha$-scale,
$\langle {\bf B}(\alpha) {\bf B}(\alpha^\prime) \rangle
 \propto e^{- D(\alpha-\alpha^\prime)}$. 

For the optimized algorithms,
the equation for $\rho$ decouples from the equation for $J$.
After the proper averages, the learning equation for this type of drift reduces to
\begin{equation}
\frac{d\rho}{d\alpha}=\rho\left\langle  W
\left(\frac{y\tilde{\sigma}_B}{\rho}-\Delta\right)-
\frac{1}{2}W^{2}\right\rangle - \rho D \:,
\end{equation}

This equation can be solved for each particular modulation function $W$. 
The generalization error $e_G=\frac{1}{\pi} \arccos(\rho)$ 
for the several algorithms described in the last
section are compared in Fig.~\ref{eg-random1} for the noiseless
case. Solid curves refer to
integrations of the above learning equation and symbols correspond to
simulation results. Although the rule 
is continuously changing, it can be tracked within an stationary error
$e_G^\infty$ which depends on the drift amplitude $D$. The functions
$e_G^\infty(D)$ for the various algorithms
can be found from the condition $d\rho/d\alpha=0$ 
and are shown in Fig.~\ref{eg-random2}.

\begin{table}[htb]
\begin{center}
\caption{Small drift exponents: Random  case}
\vspace{.5cm}
\begin{tabular}{||c|c|c||}
\hline
\centering
                       &$e_G^\infty(D) $& $e_G^\infty(D=0,\chi=0) $\\ 
\hline\hline
{\bf Annealed Hebb  } & $ \left(\frac{D}{\pi^3}\right)^{1/4}\approx
0.42\: D^{1/4} $&$0.40\: \alpha^{-1/2}$\\
\hline
{\bf Symmetric  } &$ \left(\frac{\sqrt{2}}{\pi^2}\right)^{1/3} D^{1/3}\approx 0.52 \:D^{1/3}$&$1.41\: \alpha^{-1}$ \\ \hline
{\bf Step   } &$ \frac{(4)^{1/3}}{\pi} D^{1/3} \approx 0.51
\:D^{1/3}$&$1.27\:\alpha{-1}$ \\\hline
{\bf Optimal   } &$\left(\frac{2}{\pi^2 A^2}\right)^{1/3}D^{1/3} \approx
 0.45 \:D^{1/3} $&$0.88 \:\alpha^{-1}$\\\hline
\end{tabular}
\end{center}
\label{table1}
\end{table}

This equation can be solved for each particular modulation function $W$. 
The generalization error $e_G=\frac{1}{\pi} \arccos(\rho)$ 
for the several algorithms described in the last
section are compared in Fig.~\ref{eg-random1} for the noiseless
case. Solid curves refer to
integrations of the above learning equation and symbols correspond to
simulation results. Although the rule 
is continuously changing, it can be tracked within an stationary error
$e_G^\infty$ which depends on the drift amplitude $D$. The functions
$e_G^\infty(D)$ for the various algorithms
can be found from the condition $d\rho/d\alpha=0$ and are shown in
Fig.~\ref{eg-random2}. The behavior for small drift is shown in Table
1. Note the abrupt change in the exponents due to the inclusion of
more information than the output $\tilde{\sigma}_B$.
The behavior for small drift
is shown in the Table 1. Note the abrupt change in the behavior due
to the inclusion of more information than the output $\tilde{\sigma}_B$.

\subsection{Deterministic drift}

In the learning scenario considered so far, the worst case drift will occur
when at each time step the concept is changed deterministically so that
the overlap with the current student vector is minimized. In this
situation, previously examined by Biehl and Schwarze \cite{biehl93},
the new concept is chosen by minimizing 
${\bf B}^{\mu+1}\cdot{\bf  J}^{\mu}$ subject to the conditions
\begin{eqnarray}
{\bf B}^{\mu+1}\cdot{\bf B}^\mu &=& 1- \frac{D}{N^2}  \:,  \\
{\bf B}^{\mu+1}\cdot{\bf B}^{\mu+1} &=& 1 \nonumber \:,
\end{eqnarray} 
where now $D$ is the drift amplitude for the deterministic case.
Note the different scaling with $N$ for non trivial behavior.

\begin{table}[htb]
\vspace{.5cm}
\begin{center}
\caption{Small drift exponents: Deterministic case}
\vspace{.5cm}
\begin{tabular}{||c|c|c||}
\hline
\centering
                          &$e_G^\infty(D) $&$e_G^\infty(D=0,\chi=0)$  \\ 
\hline\hline
{\bf Annealed Hebb } &
$\left(\frac{\sqrt{2D}}{\pi^2}\right)^{1/3}\approx 0.52\: D^{1/6}$&$
0.40\:\alpha^{-1/2}$ \\ \hline
{\bf Symmetric  }  &$\left( \frac{4D}{\pi^2}\right)^{1/4}\approx 0.80\: D^{1/4} $&$1.41\:\alpha^{-1}$\\ \hline
{\bf Step   }     & $\frac{2^{5/4}}{\pi} D^{1/4}\approx 0.76\:
D^{1/4}$&$1.27\:\alpha^{-1}$
\\\hline
{\bf Optimal   } & $\left(\frac{8D}{A^2}\right)^{1/4}\approx 0.63\:
D^{1/4}$
&$0.88\:\alpha^{-1}$ \\\hline
\end{tabular}
\end{center}
\label{table2}
\end{table} 
Clearly ${\bf B}^{\mu+1}$ lies in the same plane
which contains ${\bf B}^\mu$ and ${\bf J}^\mu$. 
The solution of this constrained minimization problem is
\begin{eqnarray}
{\bf B}^{\mu+1} &=& a {\bf B}^\mu - b {\bf J}^\mu \:,\\
a &=& 1-\frac{D}{N^2} + b J \rho \nonumber,\\
b & = & \frac{1}{JN} \left[\frac{2D-(D/N)^2}{1-\rho^2} \right]^{1/2} \nonumber \:.
\end{eqnarray}
In terms of $\Lambda$ and $\vec{\eta}$ of eq. 
\ref{eq:drift} is
\begin{equation}
\Lambda = (1-a)N\:, \:\:\:\: \vec{\eta} = - b {\bf J} N \:.
\end{equation}

\begin{figure}
\epsfxsize =.7\textwidth
\begin{center}
\leavevmode
\epsfbox{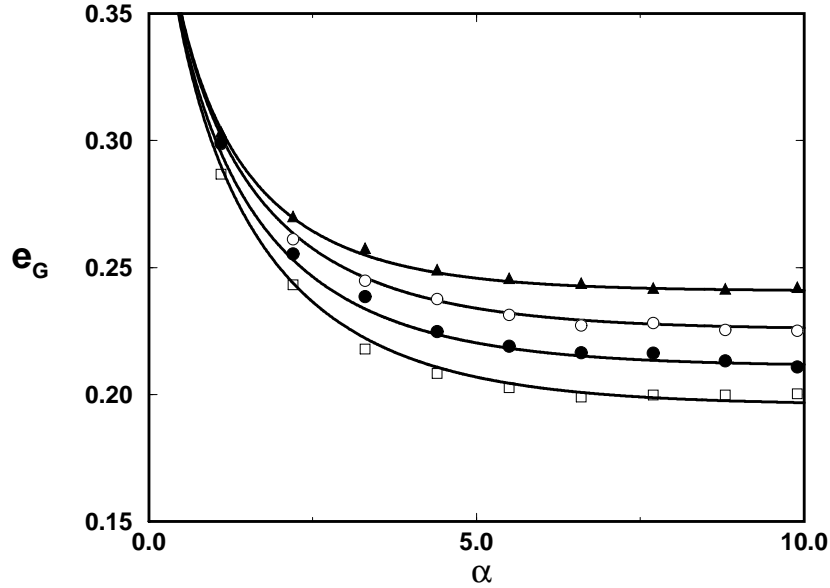}
\caption{Integration of learning equations and simulation results
($N=5000$) for deterministic drift $D=0.01$: Annealed Hebb (triangles),
Symmetric (white circles), Step (black circles) and Optimal
(white squares).} 
\label{eg-det1}
\end{center}
\end{figure}

\begin{figure}
\epsfxsize =.7\textwidth
\begin{center}
\leavevmode
\epsfbox{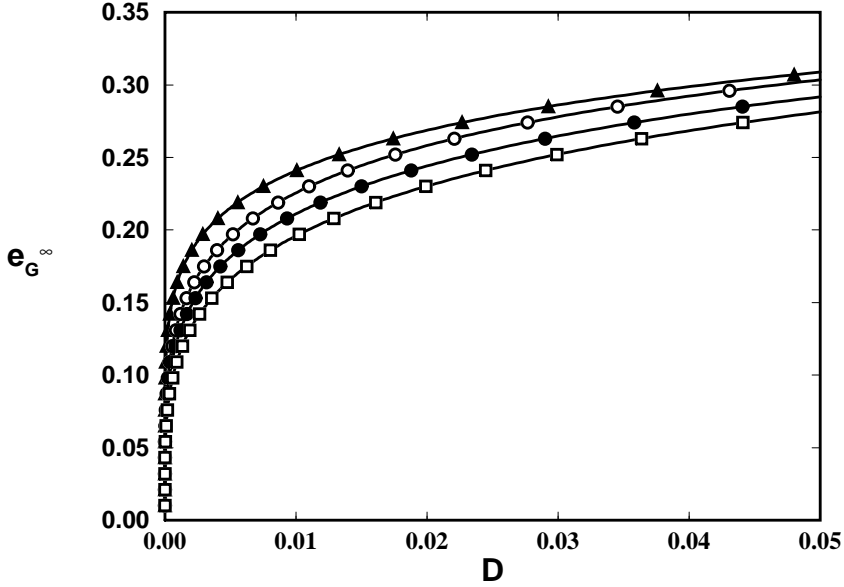}
\caption{Asymptotic error $e_G^\infty(D)$ for deterministic drift: 
Annealed Hebb (triangles),
Symmetric (white circles), Step (black circles) and Optimal
(white squares).} 
\label{eg-det2}
\end{center}
\end{figure}

The learning equation reduces to 
\begin{equation}
\frac{d\rho}{d\alpha}=\rho\left\langle  W
\left(\frac{y\tilde{\sigma}_B}{\rho}-\Delta\right)-
\frac{1}{2}W^{2}\right\rangle -\sqrt{2D(1-\rho^2)} \:,
\end{equation}
and the analysis is similar to the previous subsection. Theoretical
error curves confirmed by simulations are shown in Fig.~\ref{eg-det1}
for the different algorithms with fixed drift amplitude. The
stationary tracking error $e_G^\infty(D)$
is shown in Fig.~\ref{eg-det2}.

The behavior for small drift $D$ is shown in Table \ref{table2}. Again
we can note the occurrence of an abrupt change in the exponents after
the inclusion of information on the student's fields.

\subsection{Asymptotic behavior: Critical 
exponents and universality classes for unlearnable problems}

A simple, although partial, measure of the performance of a 
learning algorithm can be given by the asymptotic decay of the generalization 
error in the driftless case and alternatively by the residual error 
dependence on $D$ in the presence of drift. 
These are not independent aspects, but rather 
linked in a manner reminiscent of the relations between the 
different exponents that describe power law behavior
in Critical Phenomena. 

In the absence of concept drift, the generalization error decays to zero when
$\alpha$ approaches $\alpha_c$ (which here happens to be infinity but may have
a finite value in other situations \cite{watkin}) as a
power law of the number of examples with the so called {\em learning \/} or 
static exponent $\beta$,
\begin{equation}
e_G \propto \tau^{\beta} \:,
\end{equation}
where $\tau\equiv \frac{1}{\alpha} - \frac{1}{\alpha_c}$.
Thus, we may think of $e_G$ as a kind of {\em order parameter\/} in the sense
that it is a quantity which changes from zero to a finite value at $\tau=0$.
We may think of $1/\alpha$ as the analogous of the {\em control parameter\/} $T$
(temperature) in critical phenomena. 

Any amount of drift in the concept, changes the problem from a learnable
to an unlearnable one, with a residual error $e_G^\infty \equiv e_G(\tau=0)$.
We have seen that this behavior at the critical point $\tau=0$ also obeys a power law
\begin{equation}
e_G^\infty \propto D^{1/ \delta} \:,
\end{equation} 
where $\delta$ has been called the {\em drift exponent\/}.

In principle, we can classify different unlearnable situations
(due to say, various kinds of concept drift), by the two
exponents $\beta$ and $\delta$. Different algorithms and learning situations
may have the same exponents. We can thus define, in a spirit similar to 
that in the study of Critical Phenomena the so called {\em universality
classes\/} of behavior. In this paper we have seen  four classes of behavior 
$(\beta,\delta)$, the combinations $(1/2,4), (1,3)$ for the random drift
case and $ (1,4), (1/2,6)$ for the deterministic case. 
In the absence of drift $\beta=1$ for the Hebb algorithm, and $\beta=1$
for the symmetric, step and optimal algorithms we have introduced above.
There exist,
however, other classes. For example, for the standard Rosemblatt Perceptron
algorithm with fixed learning rate we have $\beta=1/3$; then, $\delta=5$ for
random drift and $\delta=8$ for deterministic drift.

We have observed that, in general, the two exponents are independent. However,
if a simple condition holds, then there exists a relation connecting $\beta$ and
$\delta$ for each kind of drift. For example, as can be compared with the above
results,
\begin{eqnarray}
\delta &=& \frac{1}{\beta} + 2 \:, \:\:\:\:\mbox{(random drift)} \label{delta}\\
\delta &=& \frac{2}{\beta} + 2 \:, \:\:\:\:\mbox{(deterministic drift)} \nonumber\:.
\end{eqnarray} 

To derive these relations
we remember that $e_G \propto \sqrt{1-\rho^2} $ when $\rho\rightarrow 1$,
so that we may write in this limit the learning equation as
\begin{equation}
\frac{de_G}{d\alpha} \approx C D^{m} e_G^{-n}- C_1(D) e_G^{n_1} - C_2(D) e_G^{n_2} - \ldots \:,
\end{equation}
where $C$ is a constant, $C_k(D)$ are functions of $D$ and $n$ and $n_k$ are positive
numbers. Now, denote by $C_*(D)$ the first function
which survives in the limit $D\rightarrow 0$, $C_*(D\rightarrow 0) = C_*$. Then, the learning equation
is

\begin{eqnarray}
\frac{de_g}{d\alpha} &\approx& - C_* e_G^{n_*} \:, \\
e_G(\alpha) &\approx & \left(  C_* (n_*-1) \alpha \right)^{-1/(n_*-1)} \:\:\:\: (n_*>1) \:,
\nonumber\\
e_G(\alpha) & \propto & e^{-C_* \alpha} \:\:\:\: (n_* = 1) \:. 
\end{eqnarray}

Thus,
\begin{equation}
\beta= 1/(n_*-1)\:, \label{beta}
\end{equation}
with $\beta \rightarrow \infty$ denoting exponential decay.

In the presence of small drift, we may write $C_1(D) \propto
D^{m_1}$. 
The stationary condition 
$de_G/d\alpha = 0$ leads to
\begin{eqnarray}
e_G^\infty(D) \approx  D^{\frac{m-m_1}{n_1+n}} \:, \\
\delta = \frac{n_1+n}{m-m_1}\:.
\end{eqnarray}
This shows that, in principle, the two exponents are
independent. However, if happens that the first surviving function 
is $C_*(D)=C_1(D)$ (which seems to be a very common situation), then 
$n_*=n_1$ and $m_1=0$, so that
\begin{equation}
\delta = \frac{1}{m}\left(\frac{1}{\beta} + (1+n) \right) \:.
\end{equation}
The relations given by Eq.~(\ref{delta}) follow from the fact that
$n=1,m=1$ for random drift and $n=0$ and $m=1/2$ for deterministic drift.
Other drift scenarios may define other universality classes.

In the case where $C_* \neq C_1$, we only can conclude that
\begin{equation}
\delta > \frac{1}{m}\left(\frac{1}{\beta} + (1+n) \right) \:.
\end{equation}
It is important to note an exponential decay of the error
$(\beta=\infty)$ leads to the limiting value $\delta =2$ both for 
deterministic as random drift. It is know that the error cannot decay 
faster than exponential in these learning problems.

\section{Practical considerations}

The most important question that can be raised in the implementation of the
variational ideas as a guide to construct algorithms is how to measure the
several unavailable quantities that go into the construction of the
modulation function. The problem of inferring the example
distribution will not be considered and only a simple method to measure the student-teacher
overlap $\rho$ will be presented. This is done by adding a `module' 
to the perceptron in order to
estimate online the generalization error, as studied in \cite{kc93}.
Algorithms that rely on this kind of module are quite robust with respect 
to changes in the distribution of examples and even to lack of
statistical independence \cite{KKC}.
Consider an online estimator (a `running average')
which uses the instantaneous error $
\epsilon ^\mu =(1-\sigma _B^\mu \sigma _J^\mu )/2$  to update the
current estimate of the generalization error:
\begin{equation}
\hat e_G^{(\mu +1)}=(1-\frac \omega N)\hat e_G^{(\mu )}+\frac \omega
N\epsilon ^\mu  \:.
\end{equation}
This estimator incorporates exponential memory loss through the $\omega $
parameter. In the perceptron, due to the factor $\lambda=\tan (\pi e_G)$ that appears in
the modulation function, fluctuations around $e_G\approx \frac 12$ may lead
to spurious divergences. Therefore it is natural to consider the truncated
Taylor expansion
\begin{equation}
\hat \lambda _k=\tan ^{(k)}(\pi \hat e_G)=\pi \hat e_G+\frac 13(\pi \hat
e_G)^3+\frac 2{15}(\pi \hat e_G)^5+\cdot \cdot \cdot +c_k(\pi \hat e_G)^k \:.
\end{equation}
Then, the modulation function for an adaptive algorithm inspired in the noiseless
optimal algorithm is 
\begin{equation}
W(\hat \lambda _k,\Delta _\mu )=\frac 1{\sqrt{2\pi }}\frac{\hat \lambda _k}{%
H(\frac{-\Delta _\mu }{\hat \lambda _k})}\exp (-\frac{\Delta _\mu ^2}{2\hat
\lambda _k^2}) \:. \label{adap}
\end{equation} 

\begin{figure}
\begin{center}
\leavevmode
\epsfxsize =.7\textwidth
\epsfbox{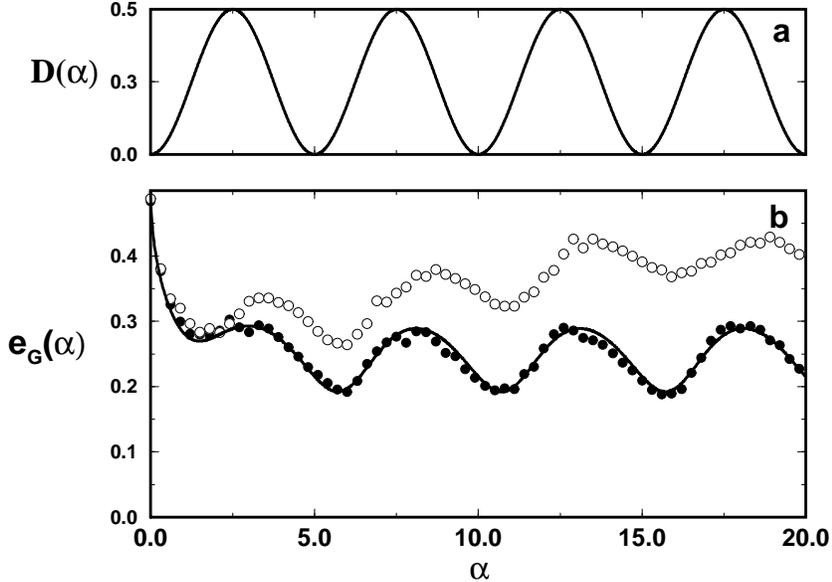}
\caption{a) Oscillating drift
level $D= D_0\sin^2(2\pi \nu t)$ for $D_0=0.5, \nu=0.1$.
b) Integration of the differential equation for the oscillating case
$D_0=0.5, \nu=0.1$ (thick solid). Adaptive optimal algorithm with 
$\omega=2$ and $k=3$ (black circles) and simple Hebb algorithm (white
circles)}
\label{accel}
\end{center}
\end{figure}

In figure \ref{accel} we present the results of applying this algorithm 
to a problem where the drift itself is non-stationary. We have dubbed this 
non-stationarity {\it drift acceleration}. The algorithm is
quite uninterested in the particular type of drift
acceleration, and as an illustration we chose a drift given by
$D= D_0\sin^2(2\pi \nu t)$. The adaptive algorithm makes {\bf no } use
of this knowledge.
There has been no attempt at optimizing
the estimator itself, but 
a reasonable and robust choice is $\omega =2$ and $k=3$. 
Simulations were done for $N=1000$, a size regime, where for all practical
purposes, the central limit theorem holds. 
Note that the Hebb algorithm is not able to keep track of the rule since
it has no internal forgetting mechanism. 

We have not studied the mixed case of drift in the presence of noise. 
The nature of the noise process corrupting the data is essential in 
determining the asymptotic learning exponent ($\beta$). While multiplicative
(flip) noise does not alter $\beta$ for the optimized algorithms, additive
(weight) noise does. This extension deserves a separate study. 
See \cite{biehl95} for the behavior of the 
optimized algorithm and noise level estimation in the presence of 
{\it noise acceleration} in the absence 
of drift, see also \cite{Heskes} where it is shown that learning is  
possible even in the mixed drift-noise case.

\section{The Wisconsin test for perceptrons: Piecewise constant rules 
\label{Wis}}

How do the algorithms studied in the previous sections perform in the case
of abrupt changes (piecewise constant rules)? The interest is in determining how do the
optimal algorithms fare in a task for which they were not optimized. The
Wisconsin test (WT) for perceptrons (WTP) to be studied here was inspired
in what is called the WT and is used in the diagnostics of pre-frontal lobe
(PFL) syndrome in human patients and which will now be described very
briefly (for details see e.g. \cite{Shallice,Levine}).

\begin{figure}
\epsfxsize =.7\textwidth
\begin{center}
\leavevmode
\epsfbox{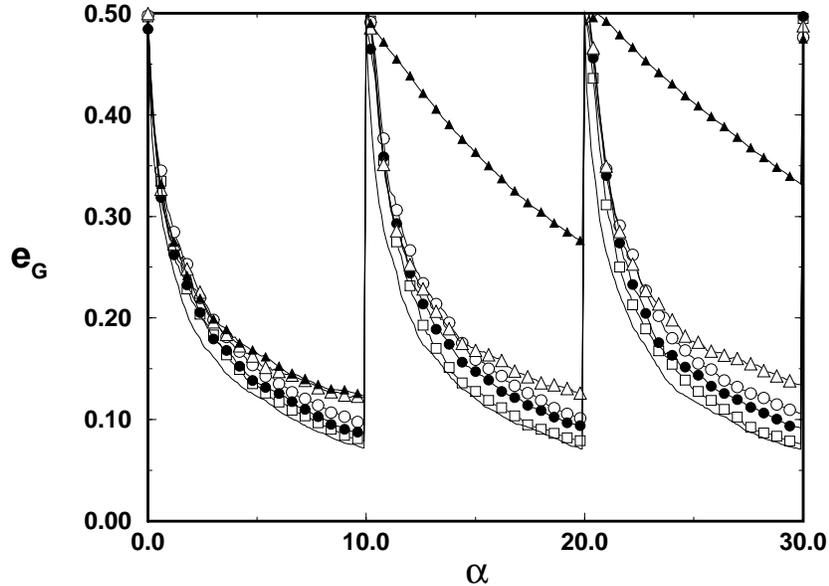} 
\caption {Simulations for $N=500$ in single runs.
Bottom line: lower bound given by the Optimal algorithm with
the true values of $\lambda$. Symbols: Optimal algorithm (white squares), 
Step algorithm (black circles), Symmetric Weight algorithm (white circles),
Annealed Hebb (white triangles), 
all with $\omega=2$ and $k=3$ estimator and and Hebb (black triangles).}
\label{fig:wisconsin}
\end{center}
\end{figure}

Consider a deck of cards, each one has a set of pictures. The cards can be
arranged in several different manners into two categories. The different
possible classifications can be done according, e.g. to color (black or red
pictures), to parity (even or odd number of figures in the picture), to
sharpness (figures can be round or pointed) etc. The examiner chooses a rule
and a patient is shown a sequence of cards and asked to classify them. The
information, whether the patient presumed classification is correct or not,
is made available before the next presentation. After a few trials (5-10)
normal subjects are able to infer the desired rule. PFL patients are
reported to infer correctly the rule after as little as 15 trials. Now a new
rule is chosen at random by the examiner but the patient is not informed
about the change. Normal patients are quick to pick up the change and after
a small number of trials (5-10) are again able to correctly classify the
cards. PFL patients are reported to persevere in the old rule and after as
much as 60 trials still insist in the old classification rule.

Our WTP is designed as a direct implementation of these ideas, by
considering learning of a piecewise constant teacher, without resetting the
couplings to {\it tabula rasa}, i.e, without letting the patient know that
the rule has changed.

Fig. (\ref{fig:wisconsin}) shows the results of simulations with the adaptive
algorithm of (\ref{adap}).
The rule is constant up to $ \alpha = 10 $, it then suddenly jumps to another, 
uncorrelated vector and stays again unchanged until $ \alpha = 20 $ and so on. 
The most striking feature is that the
perceptron with 
pure Hebbian algorithm works quite efficiently for the first rule but
perseveres in that state and adapts poorly to a change. It can not detect
performance degradation and is not surprised by the errors. 
The reason for that is that the
scale of the weight changes is the same independently of the length 
of the ${\bf J}$ vector. 
The other algorithms are able to adapt to the new conditions 
in as much as they incorporate 
the estimate of the performance of the student.  

\section{Conclusions}

The necessary ingredients for the online tracking 
of drifting concepts, adaptation to non-stationary
noise etc., emerge naturally and in an integrated way in the optimized
algorithms. These ingredients have been {\em theoretically derived\/} 
rather than  heuristically introduced. Many of the
current ideas in machine learning of changing concepts 
can be viewed as playing a role similar
to the ideal features discussed here for the perceptron. Among the
important ideas arising from the variational approach are:
\begin{itemize}
\item {\bf Learning algorithms from first principles:}
For each one of these simple learning scenarios an 
ideal and optimal learning algorithm can be found from first principles.
These optimized algorithms do not have arbitrary parameters like
learning rates, acceptance thresholds, parameters of learning schedules,
forgetting factors etc. Instead, they have a set of features which play
similar roles to those heuristic procedures. The exact form of these
features may suggest new mechanisms for better learning.
\item {\bf Learn to learn in changing environments:}
The Optimal modulation  function $W$ indeed represents a parametric
family of functions, the (non-free) parameters being 
the same as those present in the
probability distribution of the learning problem: $J$, $\rho$, $\chi$, etc.
The modulation function changes during learning, that is, the 
algorithm moves in this parametric space during the learning process. The 
student ``learns to learn'' by online estimation of the 
necessary parameters of its modulation function.
\item {\bf Robustness of optimized algorithms:} Historically, a multitude of
learning algorithms has been suggested for the perceptron: Rosemblatt's
Perceptron, Adaline, Hebb, 
Adatron, Thermal Perceptron, OLGA, etc. From the
variational perspective, these practical algorithms can 
be viewed as more or less
reliable approximations of the ideal ones in the TL. 
For example, simple Hebb corresponds to
the Optimal algorithm in the
limit $\rho\rightarrow 0$; Adatron (Relaxation) algorithm is related to 
the limit $\rho\rightarrow 1$; OLGA \cite{KS} and Thermal perceptron \cite{Frean}
includes an acceptance
threshold which mimics the optimal algorithm in the presence of
multiplicative noise $\chi$. 
Thus, although the optimal algorithms are derived for
very specific distributions of examples, it does not mean that they are 
fragile, non-robust when applied in other environments. They are indeed
very robust \cite{KKC}, at least
for the environments in which the standard algorithms work, since
these practical algorithms are `particular cases' of a more general
modulation function. But since new learning situations (new types of noise,
drifting processes, general non-stationarity etc.) can be theoretically
examined from the variational viewpoint, it is possible that new features
emerge, and that these suggest new practical ideas for more robust and
efficient learning. 
\item {\bf Emergence of `cognitive' modules:} Have the variational ideas
any relevance to `Biological Machine Learning'? Probably not for the
biological {\em structures\/}, which are
produced by opportunistic `evolution bricolage',
but perhaps they might apply in understanding
biological cognitive {\em functions\/}. 
The variational approach brings forth
a suggestion that, even if not new, acquires a more concrete form due to 
the transparent nature of the simple models studied: 
{\em optimization of the learning ability leads to
the emergence of `cognitive functional modules', here defined as components
of the modulation function and accessory 
estimators of relevant quantities of the probability distribution
related to the learning situation\/}. 
A tentative list of such
estimators suggested by the variational approach may be: a) a 
{\em mismatch\/} (surprise)
module for detection of discrepant examples; 
b) an emotional/attentional module for providing differential memory weight
for these discrepant examples; c) `constructivist' filters which accommodate 
or downplay
the highly discrepant data; d) noise level estimators for tuning these filters;
e) a working memory system for online estimation of current performance which enables
detection of environmental changes.
In conclusion, the variational approach suggests that the {\em necessity\/}
of certain {\em cognitive functions\/}
may be related to statistical 
inference principles already present in simple learning machines.
\item{\bf Extensions:} All the results here presented have been obtained under
a rather severe set of restrictions from a practical point of view. 
The main points concern the TL; noise, order parameter and example 
distribution estimation; larger architecture complexity. 
At present we don't know how to handle finite size effects. 
That the parameter estimation problem is probably easier than the others 
is suggested by the robustness found in \cite{wusp}.
The extension of the variational program 
to architectures experimentally more relevant, 
specially to include hidden nodes and soft transfer functions
\cite{vc97,saadglobal} has shown that this is a difficult task. The effects that drift
may have are not known, but it could even induce faster breaking
of the permutation symmetry among the hidden nodes, 
thus affecting the plateaus structure. 
\end{itemize}

Important remaining questions  concern whether the variational 
approach can be successfully applied to other learning models 
(radial basis functions, mixture models etc.).
The answers  will help in determining the difference between universal and
particular  features of the learning systems.

{\bf Acknowledgments:} OK and RV were supported by FAPESP fellowships. 
NC was partially supported by CNPq and FINEP/RECOPE.

\end{document}